\PassOptionsToPackage{dvipsnames,table}{xcolor}
\documentclass[unnumsec,webpdf,modern,large]{mam-authoring-template}%

\graphicspath{{Fig/}}

\usepackage[version=3]{mhchem} 
\usepackage{booktabs}
\usepackage{pdfpages}
\usepackage{placeins} 
\usepackage{amssymb}  
\usepackage{amsmath}  
\usepackage{threeparttable}
\usepackage{lmodern} 
\usepackage{nameref}

\usepackage{hyperref}
\hypersetup{
    colorlinks=true,
    citecolor=BrickRed, 
    linkcolor=BrickRed,
    filecolor=BrickRed,      
    urlcolor=BrickRed,
    pdftitle={Overleaf Example},
    pdfpagemode=FullScreen,
    }

\usepackage{todonotes}

\renewcommand{\vec}[1]{\boldsymbol{#1}}

\theoremstyle{thmstyleone}%
\theoremstyle{thmstyletwo}%
\theoremstyle{thmstylethree}%

\immediate\write18{pdflatex 02_SI.tex}  

\begin{document}

\journaltitle{arXiv}
\DOI{DOI HERE}
\copyrightyear{2025}
\pubyear{2014}
\access{Advance Access Publication Date: Day Month Year}
\appnotes{Original Article}

\firstpage{1}

\title[PtyRAD]{PtyRAD: A High-performance and Flexible Ptychographic Reconstruction Framework with Automatic Differentiation}

\author[1,$\ast$]{Chia-Hao Lee\ORCID{0000-0001-8567-5637}}
\author[2]{Steven E. Zeltmann\ORCID{0000-0003-1790-3137}}
\author[1,3]{Dasol Yoon\ORCID{0000-0003-2284-7010}}
\author[1]{Desheng Ma\ORCID{0000-0001-5237-0933}}
\author[1,4,$\ast$]{David A. Muller\ORCID{0000-0003-4129-0473}}

\authormark{Lee et al.}

\address[1]{\orgdiv{School of Applied and Engineering Physics}, \orgname{Cornell University}, \state{Ithaca, New York}, \country{USA}}
\address[2]{\orgdiv{Platform for the Accelerated Realization, Analysis, and Discovery of Interface Materials}, \orgname{Cornell University}, \state{Ithaca, New York}, \country{USA}}
\address[3]{\orgdiv{Department of Materials Science and Engineering}, \orgname{Cornell University}, \state{Ithaca, New York}, \country{USA}}
\address[4]{\orgdiv{Kavli Institute at Cornell for Nanoscale Science}, \orgname{Cornell University}, \state{Ithaca, New York}, \country{USA}}

\corresp[$\ast$]{Corresponding authors. \href{email:chia-hao.lee@cornell.edu}{chia-hao.lee@cornell.edu}, \href{email:david.a.muller@cornell.edu}{david.a.muller@cornell.edu}}

\received{Date}{0}{Year}
\revised{Date}{0}{Year}
\accepted{Date}{0}{Year}

\abstract{Electron ptychography has recently achieved unprecedented resolution, offering valuable insights across diverse material systems, including in three dimensions. However, high-quality ptychographic reconstruction is computationally expensive and time consuming, requiring a significant amount of manually tuning even for experts. Additionally, essential tools for ptychographic analysis are often scattered across multiple software packages, with some advanced features available only in costly commercial software like MATLAB. To address these challenges, we introduce PtyRAD (Ptychographic Reconstruction with Automatic Differentiation), an open-source software framework offers a comprehensive, flexible, and computationally efficient solution for electron ptychography. PtyRAD provides seamless optimization of multiple parameters---such as sample thickness, local tilts, probe positions, and mixed probe and object modes---using gradient-based methods with automatic differentiation (AD). By utilizing PyTorch’s highly optimized tensor operations, PtyRAD achieves up to a 24$\times$ speedup in reconstruction time compared to existing packages without compromising image quality. In addition, we propose a real-space depth regularization, which avoids wrap-around artifacts and can be useful for twisted two-dimensional (2D) material datasets and vertical heterostructures. Moreover, PtyRAD integrates a Bayesian optimization workflow that streamlines hyperparameter selection. We hope the open-source nature of PtyRAD will foster reproducibility and community-driven development for future advances in ptychographic imaging.}

\maketitle


\section{Introduction}\label{Introduction}

Ptychography has emerged as a high-resolution, dose-efficient phase retrieval technique by computationally reconstructing both the object and the illumination from a series of diffraction patterns (\cite{rodenburg2019ptychography}). 
Originally proposed by~\cite{Hoppe:a06516} primarily to solve the phase problem in crystallography using overlapping microdiffraction patterns, its ability to retrieve phase information has been recognized as a potential approach to overcome the lens aberrations that limit the resolution of electron microscopes (\cite{rodenburg1993experimental, nellist1995resolution, humphry2012ptychographic, putkunz2012atom}) and found widespread application for lensless imaging with X-rays (\cite{chapman2010coherent, rodenburg2007hard}). 
For a comprehensive review of the early state of the field, we refer readers to~\cite{rodenburg2019ptychography}.


Recently, electron ptychography has achieved resolution well beyond the diffraction limit, reaching sub-0.5 \AA\ in atomically thin 2D materials (\cite{jiang2018electron, nguyen2024achieving}). 
Such advances and a renewed interest in the method were driven by high-dynamic-range and high-speed direct electron detectors (\cite{tate2016high, philipp2022very, zambon2023kite, ercius20244d}) that can simultaneously resolve bright direct beams alongside weak high-angle scattering while overcoming drift and instrument instability.
Another key development is multislice electron ptychography (MEP), which adopts the multislice formalism (\cite{maiden2012ptychographic, van2013general, goddenPtychographicMicroscopeThreedimensional2014, Tsai2016multislice}) to relax the specimen thickness limitation. 
By explicitly modeling the multiple scattering and the evolution of the probe as it propagates through the specimen, MEP achieves lateral resolution primarily limited by atomic vibrations rather than the imaging system (\cite{chen2021electron}), while also enabling depth-resolved information to be extracted from a single scan (\cite{gao2017electron, chen2021electron, dongVisualizationOxygenVacancies2024}). 
Most importantly, such three-dimensional (3D) reconstructions are less susceptible to channeling artifacts, which often hinder defect localization in thick crystals when using other optical sectioning techniques with annular darkfield (\cite{kimoto2010local}), annular brightfield (\cite{gao2018picometer}), or differential phase contrast (\cite{close2015towards}) imaging.

Reconstruction algorithms are essential parts of ptychography and can be broadly classified into direct and iterative approaches.
Direct methods provide fast reconstructions for live imaging and beam sensitive applications (\cite{strauch2021live, o2020phase}), but these methods do not incorporate scattering outside the objective aperture so the resolution is limited to twice the illumination angle, nor do they correct for channeling in the the sample (\cite{rodenburg1992theory, rodenburg1993experimental, pennycook2015efficient}).
Recently, machine-learning-based techniques have been explored as alternative direct solvers (\cite{cherukara2020ai, friedrich2022phase, chang2023deep}).
While these methods offer significant speed advantages over iterative approaches, their performance is inherently tied to the training data, often limiting their applicability to a narrower range of experimental conditions and sample types.

In contrast, iterative algorithms refine the object and probe estimates through successive steps until the model agrees with the measured diffraction data. 
This approach provides the flexibility to incorporate more complex scattering processes and enables the reconstruction of details beyond the diffraction limit.
Various iterative ptychographic algorithms are inspired by or adapted from classical phase retrieval methods (\cite{gerchberg1972practical, fienup1978reconstruction, fienup1982phase, luke2004relaxed, elser2003phase}), with the extended ptychographic iterative engine (ePIE) (\cite{maiden2009improved}) and the gradient-based least-squares maximum likelihood framework (LSQML) (\cite{odstrvcil2018iterative}) emerging as the most widely used approaches today.

One particular advantage of iterative methods is their ability to incorporate essential features for modeling experimental imperfections and complex physical models---such as position correction (\cite{maiden2012annealing}), mixed-state probe retrieval (\cite{thibault2009probe, thibault2013reconstructing, odstrcil2016ptychographic}), and multislice object modeling (\cite{Tsai2016multislice}).
These capabilities have made iterative approaches the dominant choice for high-resolution ptychography, and several open-source iterative ptychography packages (\cite{enders2016computational, Wakonig2020PtychoShelves, jiang2020, Savitzky2021py4DSTEM, loetgering2023ptylab, varnavides2023iterative}) have been developed, each implementing different features and tailored to specific use cases. 
A more comprehensive list of open-source ptychography packages is given in SI Table S1.

Despite their strengths, a key limitation of the existing iterative packages is that efficient reconstructions are only possible when analytical update rules are available for the forward model. 
Introducing new optimizable parameters often require either manually deriving the update steps, estimating the gradients in a computationally expensive finite-difference manner, or refining them as hyperparameters for the entire reconstruction task. 
Therefore, many recent ptychography implementations (\cite{kandel2019using, du2020three, du2021adorym, seifert2021efficient, guzzi2022modular, diederichs2024exact, wu2024dose}), primarily developed in the X-ray community, have adopted automatic differentiation (AD) (\cite{rall1981automatic}) for gradient computation.
The AD algorithm systematically applies the chain rule to calculate gradients from any composition of differentiable operations, making it an ideal framework for rapid prototyping new forward models or building on existing models.

Here, we introduce PtyRAD (Ptychographic Reconstruction with Automatic Differentiation), an iterative gradient descent ptychography package that harnesses AD to provide a flexible and efficient approach to ptychographic reconstructions. PtyRAD is an open-source framework developed with the following guiding principles:

\begin{enumerate}
  \item Accessibility: Fully implemented in Python, with comprehensive documentation.
  
  \item Performance: Efficient reconstruction pipelines, supporting single and multi-GPU acceleration on various platforms.
  
  \item Flexibility: A modular architecture allowing easy customization of forward models, optimization schemes, and physical constraints.

  \item Comprehensive Workflow: A complete suite of tools covering preprocessing, reconstruction, and hyperparameter tuning, streamlining the entire reconstruction pipeline.
\end{enumerate}

We first briefly discuss the major components of PtyRAD and our implementation of the mixed-state multislice ptychography model.
We then provide detailed benchmarks on published datasets, comparing computational efficiency and reconstruction quality with other commonly-used electron ptychography packages.
We further demonstrate how incorporating constraints and regularizations improves the reconstruction quality, including an enhanced depth regularization approach for non-periodic structures.
Additionally, we introduce a fast and effective method for hyperparameter selection.
Lastly, we summarize key contributions and provide potential future directions.

\begin{figure*}[ht]%
\centering
\includegraphics[width=0.95\linewidth]{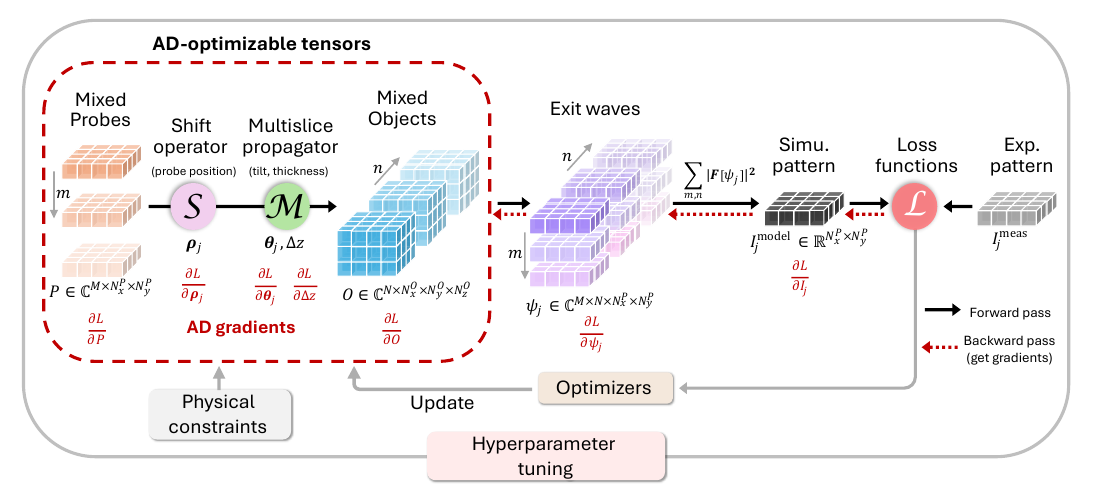}
    \caption{Schematic of the PtyRAD framework for ptychographic reconstruction using automatic differentiation (AD). The forward model simulates ptychographic measurements at each scan position $\vec{\rho}_j$ by first applying the shift operator to the probe ($S_jP = P_j$), and then propagate the shifted probe $P_j$ through the multislice object patch $O_j$ using the adaptive multislice propagator $\mathcal{M}$, parameterized by local tilt ($\vec{\theta}_j$) and slice thickness ($\Delta z$). $j$ is the labeling index used to distinguish different probe positions and associating diffraction patterns. The probe and object are both represented in mixed-state formulation with $M$ probe modes and $N$ object modes, resulting in a set of 4D mixed-state exit waves $\psi_j \in \mathbb{C}^{M \times N \times N^P_x \times N^P_y}$, which are subsequently reduced to a simulated diffraction pattern $I_j^{\text{model}} \in \mathbb{R}^{N^P_x \times N^P_y}$ by incoherently summing all states. The simulated pattern is then compared with the experimental pattern $I_j^{\text{meas}}$ through loss functions $\mathcal{L}$. PtyRAD leverages AD to compute gradients (marked in red text) via the backward pass across all participating tensors, enabling the simultaneous optimization of mixed probes, mixed objects, probe positions, tilts, and thickness using advanced PyTorch optimizers while enforcing physical constraints at each iteration. Additionally, PtyRAD seamlessly integrates a hyperparameter tuning workflow to search for promising initial parameters for each reconstruction trial.}
\label{fig_overview}
\end{figure*}

\section{Implementation of PtyRAD}\label{Implementation of PtyRAD}
PtyRAD is implemented in Python, with its core AD engine powered by PyTorch (\cite{paszke2019pytorch}), leveraging efficient GPU-accelerated tensor operations. Figure~\ref{fig_overview} illustrates major components of PtyRAD, including a unified forward model supporting mixed probe and object states with multislice; AD-powered gradient-based optimization of probe positions, object misorientations, and slice thickness; and integrated physical constraints to regularize reconstructions. PtyRAD provides an end-to-end pipeline, from initial data preprocessing to final reconstruction with efficient hyperparameter tuning, enabling researchers to rapidly prototype new algorithms, benchmark existing approaches, and foster further developments. 
See SI Figure S1 for a more detailed workflow diagram including the hyperparameter turning.

\subsection{Physical model}\label{subsec_Physical model}
PtyRAD implements the multislice (\cite{Tsai2016multislice}) and mixed-state ptychography formalisms (\cite{thibault2013reconstructing}) with an adaptive multislice propagator (\cite{Kirkland2010advanced, sha2022deep}). 
The forward model $f$ describes the physical process of how the probe $P$, object $O$, probe position $\vec{\rho}_j$, object misorientation $\vec{\theta}_j$, and slice thickness $\Delta z$\footnote{While $\Delta z$ is often colloquially referred as ``slice thickness,'' the multislice approach treats the slices as infinitely thin and $\Delta z$ is the separation distance between slices.} interact with each other to generate the modeled diffraction intensities $I_j^{\text{model}}$. 
\begin{equation}\label{eqn_forward}
\begin{aligned}
    I^{\text{model}}_j(\vec{k}) &=f(P, \, O, \, \vec{\rho}_j,\, \vec{\theta}_j, \Delta z)\\
\end{aligned}
\end{equation}

The index $j$ is used to label each probe position and the associated diffraction pattern. The variables carrying the subscript can have distinct values for each index. $j$ ranges from 1 to $N_{tot}$, where $N_{tot}$ is the total number of probe positions. All of the model parameters in Equation~\ref{eqn_forward} are AD-optimizable. 

Following the mixed state formalism (\cite{thibault2013reconstructing}), the probe $P$ and object $O$ arrays are represented as mixtures of mutually incoherent states (Figure~\ref{fig_overview}).
Specifically, the probe is given by $P \in \mathbb{C}^{M \times N^P_x \times N^P_y}$
, with individual states denoted as $P^{(1)}, P^{(2)}, \dots, P^{(M)}$ for $M$ probe modes.
Similarly, the object is represented as $O \in \mathbb{C}^{N \times N^O_x \times N^O_y \times N^O_z}$, with its modes denoted as $O^{(1)}, O^{(2)}, \dots, O^{(N)}$ for $N$ object states. 
Here, $N_x$, $N_y$, and $N_z$ along with their superscripts, denote the number of pixels along each dimension for the probe and object arrays\footnote{$O$ and $P$ may have different sizes, as the size of $P$ is most conveniently fixed by the size of the diffraction patterns, while the size of $O$ is governed by the field of view of the scan.}.
Since each state is assumed incoherent with the others, the total diffraction intensity for pattern $j$, denoted as $I_j^{\text{model}}$, is simply the summation of diffraction intensities produced from each probe and object mode combination $I_j^{(m)(n)}$. Individual intensity contribution $I_j^{(m)(n)}$ is modeled by the squared modulus of the Fourier-transformed exit waves $|\mathcal{F}[\psi_j^{(m)(n)}]|^2$, where $m \in \{1,\ldots,M\}$ and $n \in \{1,\ldots,N\}$.
\begin{equation}\label{eqn_mixed_state}
\begin{aligned}
    I^{\text{model}}_j(\vec{k})
    &= \sum_{m=1}^{M} \sum_{n=1}^{N} I_j^{(m)(n)}\\
    &=\sum_{m=1}^{M} \sum_{n=1}^{N} |\mathcal{F}[\psi_j^{(m)(n)}(\vec{r})]|^2
\end{aligned}
\end{equation}
The Fourier transform ($\mathcal{F}$) describes the far-field diffraction as the real-space exit wave propagates to the electron detector, which records the diffraction pattern in reciprocal space. $\vec{r} = (r_x,r_y)$ and $\vec{k} = (k_x,k_y)$ represent the real and reciprocal space coordinates, respectively.

Each exit wave $\psi_j^{(m)(n)}$ is computed using the standard multislice algorithm detailed in Chapter 6 of \cite{Kirkland2010advanced}.
For each probe position $\vec{\rho}_j$, an object patch $O^{(n)}_j$, centered on $\vec{\rho}_j$, is cropped from the full object array $O^{(n)}$.
Since object cropping is limited to integer pixel positions, we apply a shift operator to account for sub-px shifts of probe relative to the object patch.
This is expressed as $S_{\Delta \vec{r}_j}P^{(m)} = P_j^{(m)}$, where $S_{\Delta \vec{r}_j}$ is the shift operator given $\Delta \vec{r}_j$, and $\vec{\rho}_j - \textit{round}(\vec{\rho}_j) = \Delta \vec{r}_j$ represents the sub-px shift\footnote{Position correction is achieved by optimizing $\Delta \vec{r}_j$ through the shift operator because rounding and cropping operations are not differentiable.}. 
This ensures that $O^{(n)}_j$ and $P^{(m)}_j$ share the same real-space sampling and lateral extent of $N^P_x \times N^P_y$, while $O^{(n)}_j \in \mathbb{C}^{N^P_x \times N^P_y \times N^O_z}$ has an additional depth dimension with $N^O_z$ slices.
Here, $O^{(n)}_{j,1}$ denotes the first slice of the object, while $O^{(n)}_{j,N^O_z}$ denotes the last slice.

The multislice calculation is then done by sequentially transmitting through object slices $O^{(n)}_{j,1}, O^{(n)}_{j,2}, \cdots O^{(n)}_{j,N^O_z}$, while propagating with the multislice propagator $\mathcal{M}_{\vec{\theta}_j, \Delta z}$ between each slice. 
\begin{equation}\label{eqn_exit}
   \psi_j^{(m)(n)} = O^{(n)}_{j,N^O_z} \cdots \mathcal{F}^{-1}\left[ \mathcal{M}_{\vec{\theta}_j, \Delta z} \mathcal{F}\left[O^{(n)}_{j,1} P^{(m)}_j\right] \right]
\end{equation}
The multislice propagator depends on the slice thickness $\Delta z$ and includes the local mistilt $\vec{\theta}_j = (\theta_{j,x}, \theta_{j,y})$, following equation 6.99 in \cite{Kirkland2010advanced}. 
\begin{equation}
\mathcal{M}_{\vec{\theta}_j, \Delta z}(\vec{k}) = \exp \left[ -i\pi \lambda |\vec{k}|^2 \Delta z + 2\pi i \Delta z (k_x\tan \theta_{j,x} + k_y \tan \theta_{j,y}) \right]
\label{eqn_adaptive_propagator}
\end{equation}

\subsection{Parameter specification}
Reconstruction parameters of PtyRAD are defined via a configuration file, which can be provided in .py, .yaml, or .json formats, or reconstructions can be run interactively via a Jupyter notebook. This flexibility accommodates both scripted and exploratory workflows, allowing users to easily modify reconstruction settings and experiment with different configurations while ensuring reproducibility.

\subsection{Data import}
PtyRAD supports raw datasets with multiple formats, including .raw, .hdf5, .mat, .tif, and .npy, ensuring broad compatibility with different experimental setups. Additionally, reconstructed models (object, probe, and probe positions) from other ptychographic reconstruction packages, such as PtychoShelves/\texttt{fold\_slice} (\cite{Wakonig2020PtychoShelves, jiang2020}) and py4DSTEM (\cite{Savitzky2021py4DSTEM, varnavides2023iterative}), can be seamlessly imported as well. This feature facilitates cross-comparisons and iterative refinement, making PtyRAD suitable for integrating with existing workflows.

\subsection{Preprocessing and initialization}
PtyRAD provides a wide variety of preprocessing functions for the imported 4D-STEM data, such as permutation, reshaping, flipping, transposing, cropping, padding, and resampling. Particularly, the padding and resampling can be done in an ``on-the-fly'' manner, which greatly reduce the required GPU memory. Additionally, PtyRAD also implements features to conveniently apply Poisson noise, detector blur, and partial spatial coherence on perfect simulated data to quickly explore different experimental conditions without generating and storing redundant datasets. By default, the object is initialized with unit amplitude and small random phase perturbations uniformly sampled from the range $[0, 10^{-8}]$. This produces an object with flat amplitude and near-zero phase, which serves as a neutral starting point for optimization.

\subsection{PyTorch model and AD optimization}
Once the data is prepared, PtyRAD constructs a PyTorch model and imports the data as PyTorch tensors. We choose PyTorch because it provides a fully integrated framework with GPU acceleration, automatic differentiation (AD), a wide range of optimization algorithms, and an extensive toolkit for data processing, all backed by a large and active community. 
Its ease of use and flexibility have led to its wide adoption in both machine learning and scientific computing, enabling rapid development and excellent extensibility. 

A key motivation for the choice of PyTorch for ptychographic reconstruction is its support for automatic differentiation (AD) (\cite{rall1981automatic}), which efficiently computes exact gradients by decomposing complex operations into elementary components and systematically applying the chain rule, making AD the backbone of backpropagation in modern machine learning. 
AD allows PtyRAD to flexibly incorporate sophisticated physics models, such as multiple scattering and partial coherence in a simple and unified way. Unlike conventional methods that require manually deriving update steps, AD eliminates this burden entirely: one only needs to define the forward model and the loss function, and all gradient computations follow automatically. This makes AD an ideal framework for incorporating new optimizable parameters and rapidly prototyping novel reconstruction algorithms. 

\subsection{AD-optimizable parameters}
Currently, PtyRAD implements six AD-optimizable parameters (Equation~\ref{eqn_forward}), including object amplitude, object phase, object tilt, probe, probe position, and slice thickness. Note that PtyRAD represents the complex object function using two independent real-valued arrays for amplitude and phase, which are recombined into a complex transmission function during the forward pass, while allowing their learning rates to be specified separately. This design allows independent control over the learning rate and onset---i.e., the iteration at which specific parameters begin optimization---enabling precise tuning of the reconstruction process. For example, one may fix the object amplitude at 1 for a pure phase object approximation or delay probe and position optimization until a rough object structure is retrieved. To improve convergence and stability, it is often beneficial to introduce optimizable parameters gradually rather than all at once. Since the computational cost of AD scales with the number of active parameters, starting with a smaller set and progressively refining the reconstruction in a pyramidal approach balances efficiency and accuracy.

\subsection{Optimizers and loss functions}
PtyRAD supports all 14 gradient-based optimization algorithms provided by PyTorch, with Adam (\cite{kingma2014adam}) as the default choice due to its adaptive learning rate and robustness in handling noisy gradients.

In addition to optimizer selection, PtyRAD permits flexibility in the construction of the loss function. 
We primarily utilize the negative log likelihood functions, $\mathcal{L}_{\text{Gaussian}}$ and $\mathcal{L}_{\text{Poisson}}$, which are adapted from the original derivation in \cite{thibault2012maximum}, assuming Gaussian or Poisson noise statistics for the measured data, respectively.

\begin{equation}
\mathcal{L}_{\text{Gaussian}} = 
\frac{\sqrt{\left\langle \left( I_{\text{model}}^p - I_{\text{meas}}^p \right)^2 \right\rangle_{\mathcal{D,B}}}}
{\langle I_{\text{meas}}^p \rangle_{\mathcal{D,B}}}
\end{equation}
\begin{equation}
\mathcal{L}_{\text{Poisson}} = 
- \frac{\left\langle I_{\text{meas}}^p \log(I_{\text{model}}^p + \epsilon) - I_{\text{model}}^p \right\rangle_{\mathcal{D,B}}}
{\langle I_{\text{meas}}^p \rangle_{\mathcal{D,B}}}
\end{equation}
Here, $I_{\text{model}}$ and $I_{\text{meas}}$ denote the modeled and measured diffraction pattern. 
The patterns are raised to a chosen power $p$ when computing the loss, with default choices of $p=0.5$ for $\mathcal{L}_{\text{Gaussian}}$ and $p=1$ for $\mathcal{L}_{\text{Poisson}}$. 
The $\langle \cdot \rangle_S$ represents averaging over a certain dimension $\mathcal{S}$, where $\mathcal{D}$ denotes the detector dimension, and $\mathcal{B}$ denotes the batch\footnote{Iterative ptychographic reconstructions often group diffraction patterns as $I_B \in \mathbb{R}^{B \times N_x\times N_y}$ for efficient GPU utilization. $B$ is often called the ``batch size.''}, and $\mathcal{R}$ denotes the spatial dimension. 
We added $\epsilon = 10^{-6}$ into the $\mathcal{L}_{\text{Poisson}}$ formula for numerical stability.

Additionally, we have implemented a PACBED loss, which promotes consistency with the Position-Averaged Convergent Beam Electron Diffraction (PACBED) pattern (\cite{lebeau2010position}), and an explicit sparsity-promoting regularization $\mathcal{L}_{\text{sparse}}$ containing the $L^p$-norm of the object phase (\cite{candes2006near, tibshirani1996regression}).
By default, we use $p=1$, corresponding to L1 regularization.
We are also interested in extending PtyRAD with L0-type sparsity constraints using Fourier-domain thresholding for low dose ptychography (\cite{moshtaghpour2025lorepie}) in future versions.

\begin{equation}
\mathcal{L}_{\text{PACBED}} = 
\frac{\sqrt{
\left\langle \left( \langle I_{\text{model}} \rangle_{\mathcal{B}}^p - \left\langle I_{\text{meas}} \right\rangle_{\mathcal{B}}^p \right)^2 \right\rangle_{\mathcal{D}}}}
{\langle I_{\text{meas}}^p \rangle_{\mathcal{D,B}}}
\end{equation}
\begin{equation}\label{eqn_sparse}
\mathcal{L}_{\text{sparse}} = 
\left\langle \left| O_p \right|^{p} \right\rangle_{\mathcal{R,B}}^{\frac{1}{p}}
\end{equation}

The objective function for optimization is given by a weighted combination of the above terms, with user-controlled weights $w$
\begin{equation}\label{eqn_losses}
\mathcal{L}_{\text{total}} = w_1\mathcal{L}_{\text{Gaussian}} + w_2\mathcal{L}_{\text{Poisson}} + w_3\mathcal{L}_{\text{PACBED}} + w_4\mathcal{L}_{\text{sparse}}
\end{equation}

\subsection{Physical constraints}
Inverse imaging problems are inherently ill-posed, making it essential to incorporate physical constraints to help regularize the optimization process, mitigate artifacts, and enforce physically meaningful reconstructions. As shown in Figure~\ref{fig_overview}, these constraints are applied to the probe, object, and tilt arrays at each iteration after taking the gradient descent step. PtyRAD provides a diverse set of physical constraints, including orthogonalization of the mixed-state probes (\cite{thibault2013reconstructing}); cutoff mask to constrain the probe in Fourier space; object blurring in real and reciprocal space to ensure stability in multislice reconstruction; and a series of thresholding, positivity, and complex relation (\cite{clark2010use, xu2024high}) that regularize the object amplitude and phase to enforce physically reasonable solutions with improved interpretability, particularly in multislice reconstructions. Our framework allows each constraint to be applied at set intervals during reconstruction and with relaxation to permit partial deviation from the regularization condition. 

\subsection{Hyperparameter tuning}
Convergence in iterative optimization problems is highly sensitive to algorithmic parameters such as batch sizes, learning rates, and other configurational settings which are generally referred to as hyperparameters (\cite{bergstra2011algorithms}). Automatic parameter selection based on Bayesian optimization (\cite{cao2022automatic}) has been shown to provide faster and more optimal choice of these parameters as compared to hand-tuning by human experts. Moreover, experimental parameters such as specimen thickness and crystal tilts are often imprecisely known, while others---such as the semi-convergence angle, probe aberrations, and scan distortion---are challenging to optimize via gradient descent simultaneously with the object reconstruction. Bayesian optimization provides a better balance of exploration and exploitation for parameters that have many local minima or complex coupling to the object reconstruction trajectory, and so has been widely implemented for optimization of this class of parameters. PtyRAD uses Optuna (\cite{akiba2019optuna}), a versatile hyperparameter tuning framework that offers a wide range of algorithms beyond Bayesian optimization. In addition, Optuna provides efficient pruning algorithm and distributed optimization capability to search the parameter space even more efficiently.
The high-level and non-AD-optimizable parameters of the reconstruction are refined by the hyperparameter tuning process which encloses the AD reconstruction loop.
SI Table S2 summarizes the tunable hyperparameters currently implemented in PtyRAD, including batch size, learning rates, defocus, and more.

\section{Results and Discussion}\label{Results and Discussion}
In order to demonstrate the performance of PtyRAD, we have performed a series of benchmarking, speed, and convergence tests using publicly available ptychography datasets. 
We then  explore the impact of the regularizations and physical constraints available for improving reconstruction quality as well as the performance of the hyperparameter tuning workflow for rapidly determining optimal reconstruction parameters.

\begin{figure*}[ht]%
\centering
\includegraphics[width=0.9\linewidth]{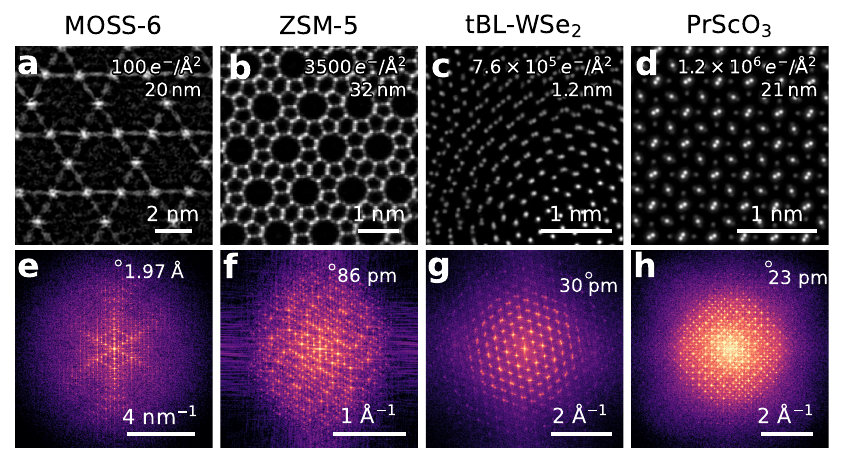}
\caption{(a–d) Ptychographic phase images reconstructed using PtyRAD. These examples are from publicly available datasets acquired under various experimental conditions and material systems, including a metal-organic framework (MOSS-6), a zeolite (ZSM-5), a twisted bilayer transition metal dichalcogenide (tBL-\ce{WSe2}), and a rare-earth oxide (\ce{PrScO3}). The sample thickness and electron doses are indicated in the top right corner of each panel. (e–h) Corresponding fast Fourier transform (FFT) power spectra of (a–d), demonstrating high information transfer of the reconstructed images. These results highlight PtyRAD’s capability to successfully reconstruct datasets across a wide range of doses and specimen thicknesses.}\label{fig_exp_examples}
\end{figure*}

\subsection{Benchmarking on published datasets}
In order to benchmark the performance of PtyRAD, we performed multislice ptychography reconstructions from a selection of publicly-available datasets from the literature (\cite{liAtomicallyResolvedImaging2025, zhangThreedimensionalInhomogeneityZeolite2023, nguyen2024achieving, chen2021electron}, spanning a wide range of electron doses and sample thicknesses. The resulting depth-summed phase images are shown in Figure~\ref{fig_exp_examples}a--d with their corresponding fast Fourier transform (FFT) power spectra in Figures ~\ref{fig_exp_examples}e--h. 
See SI Table S3 for the major experimental and reconstruction parameters.
Full details on the experimental conditions for each dataset are available in the original references, and PtyRAD input files to reproduce these reconstructions are provided in our Zenodo record.

To demonstrate reconstruction of very low-dose data from radiation sensitive materials, we show in Figure~\ref{fig_exp_examples}a a reconstruction of the dataset taken from a metal-organic framework MOSS-6 at 100~e$^-$/\AA$^2$ reported by~\cite{liAtomicallyResolvedImaging2025}. At such low electron dose, the resolution is essentially dose-limited, and we observe the same information limit as reported in the original paper and are similarly able to resolve individual atoms in the linker nodes. The ability to reconstruct this dataset demonstrates the robustness of the AD approach to experimental data with severe shot noise. A dataset taken from the zeolite ZSM-5 at a higher electron dose of 3500~e$^-$/\AA$^2$ is shown in Figure~\ref{fig_exp_examples}b, using the original data taken from~\cite{zhangThreedimensionalInhomogeneityZeolite2023}. Here the resolution is also dose-limited, but the higher dose permits an information limit of 86~pm, consistent with the resolution reported in the original work. 

We also demonstrate sub-0.5 \AA\ resolution reconstructions from radiation-hard samples imaged at high electron doses. Figure~\ref{fig_exp_examples}c shows a twisted bilayer of~\ce{WSe2} using the dataset from~\cite{nguyen2024achieving} at a dose of $7.6 \times10^{5}$ e$^-$/\AA$^2$ and using a probe-corrected instrument. Using PtyRAD for the multislice electron ptychographic reconstruction, we observe an information limit of 30~pm, superior to the 41~pm limit reported by the original work using the single slice model. Finally, Fig~\ref{fig_exp_examples}d shows a reconstruction of the~\ce{PrScO3} dataset reported by \cite{chen2021electron} at a dose of approximately 10$^6$~e$^-$/\AA$^2$. Using PtyRAD, we achieve good information transfer down to 23~pm, similar to that reported in the original work. That work attributed the observed width of the atomic columns primarily to random thermal displacement of the atoms at room temperature, rather than limitation of the imaging method, and estimated an instrument blur of approximately 16~pm.

These four benchmarking reconstructions, which cover samples from bilayer to 32~nm in thickness and span four orders of magnitude of dose, demonstrate the applicability of PtyRAD and AD-based ptychography to a wide range of relevant experimental parameters. 
We successfully reconstruct data ranging from sparse, low dose patterns dominated by shot noise, up to data with very high signal to noise, strong dynamical scattering effects, and substantial darkfield information. 
While thorough comparisons of reconstructions on experimental data are difficult, we show high-quality phase images with equivalent or better information limit in each example.

\begin{table*}[ht]
\begin{threeparttable}
\centering
\caption{Feature comparison of different iterative ptychography packages.\label{table_features}}
\tabcolsep=7pt
\begin{tabular}{c c c c c c c c c c}
\toprule
\rowcolor[HTML]{EED5D7}
\textbf{Package} & \begin{tabular}{@{}c@{}}\textbf{Multi} \\ \textbf{slice}\end{tabular} & \begin{tabular}{@{}c@{}}\textbf{Mixed} \\ \textbf{Probe}\end{tabular} & \begin{tabular}{@{}c@{}}\textbf{Mixed} \\ \textbf{Object}\end{tabular} & \begin{tabular}{@{}c@{}}\textbf{Position} \\ \textbf{Correction}\end{tabular} & \begin{tabular}{@{}c@{}}\textbf{Tilt} \\ \textbf{Correction}\end{tabular} & \begin{tabular}{@{}c@{}}\textbf{Ptycho} \\ \textbf{Tomo}\end{tabular} & \begin{tabular}{@{}c@{}}\textbf{Hyperparam} \\ \textbf{Tuning}\end{tabular} & \textbf{Language} & \textbf{GPU} \\
\midrule
\rowcolor[HTML]{F6ECEC}
PtyRAD (ours)          & \checkmark          & \checkmark           & \checkmark            & \checkmark                   & \checkmark\tnote{c}       &        & \checkmark                 & Python           & Multi         \\
\rowcolor[HTML]{F6ECEC}
PtychoShelves\tnote{a}    & \checkmark          & \checkmark           &                       & \checkmark                   & Fixed value\tnote{b}   & \checkmark          & \checkmark                 & MATLAB           & Single        \\
\rowcolor[HTML]{F6ECEC}
py4DSTEM         & \checkmark          & \checkmark           &                       & \checkmark                   & Fixed value              & \checkmark       & \checkmark          & Python           & Single        \\
\bottomrule
\end{tabular}
\begin{tablenotes}
    \item[a] Refers to \texttt{fold\_slice}, the modified electron ptychography version maintained by Dr. Yi Jiang (\cite{jiang2020}).
    \item[b] This feature is implemented by another GitHub fork (variant) of \texttt{fold\_slice} maintained by LeBeau group (\cite{LeBeau2022}).
    \item[c] PtyRAD can correct either global or position-dependent tilts during the ptychographic reconstruction.
\end{tablenotes}
\end{threeparttable}
\end{table*}

\subsection{Feature comparison with other open-source packages}
To clarify the unique capabilities of PtyRAD, we compare PtyRAD with two widely used open-source packages, PtychoShelves/\texttt{fold\_slice} (\cite{Wakonig2020PtychoShelves, jiang2020}) and py4DSTEM (\cite{Savitzky2021py4DSTEM, varnavides2023iterative}) in Table~\ref{table_features}, which we select based on their current wide adoption in electron ptychography research. 
PtychoShelves originated from the X-ray ptychography community and has been extended by multiple groups to implement additional features relevant for electron ptychography, including the \texttt{fold\_slice} fork (\cite{jiang2020}) and its descendants (\cite{LeBeau2022}). 
It provides a powerful toolbox for ptychographic reconstruction, particularly for X-ray applications, supporting advanced features such as fly-scan, ptycho-tomography, and various ptychographic algorithms including difference map and ePIE. 
Py4DSTEM is a full-featured 4D-STEM analysis suite providing comprehensive tools from preprocessing to numerous virtual imaging techniques and supporting a wide range of structural characterization methods (\cite{Savitzky2021py4DSTEM,ophusAutomatedCrystalOrientation2022,donohue20214d}). 
The phase contrast imaging tools in py4DSTEM include DPC, direct ptychography, mixed-state multislice ptychography using gradient descent and other projection-based phase retrieval algorithms, as well as ptycho-tomography and novel algorithms for recovering the magnetic vector potential. 
In the following sections, we compare the performance of these packages against PtyRAD using their gradient-descent mixed-state multislice implementations. 
A more comprehensive list of available open-source ptychography packages, covering a wider range of algorithms, is provided in SI Table~S1. 
It is important to note that the information in Table~\ref{table_features} reflects the current state of these packages, but as open-source projects under current development, they are continuously evolving, and new features may be integrated over time.

All three packages share essential functionalities, including multislice algorithms (\cite{Tsai2016multislice}), mixed probe modes (\cite{thibault2013reconstructing}), and probe position correction (\cite{maiden2012annealing, thibault2012maximum}). PtyRAD expands on these capabilities by introducing support for mixed states in the object function. Reconstruction of mixed object states have been demonstrated in optical ptychography (\cite{thibault2013reconstructing, li2016breaking}), and recent studies have investigated its potential for reconstructing phonon modes (\cite{gladyshev2023reconstructing}) and improving reconstruction quality (\cite{schloz2024improved}) in electron ptychography.
We note that while PtyRAD includes full support for mixed object states, all the reconstructions shown in this manuscript use a single object state. A thorough exploration of the advantages and implications of mixed objects will be presented in a follow-up study.

Taking advantage of automatic differentiation (AD), PtyRAD also introduces position-dependent specimen tilt correction, with the option to switch between local and global tilt corrections. This feature addresses practical challenges in experiments, as unavoidable mistilts or bending in the specimen can impact reconstruction accuracy. Local tilt correction is particularly valuable for large field-of-view datasets (\cite{kp2025microscopicmechanismsflexoelectricityoxide}), where a global tilt correction may be insufficient. Although~\cite{sha2022deep} have reported iterative tilt and thickness correction, their implementation is not freely available. PtychoShelves and py4DSTEM support fixed global tilt corrections which are not updated during iteration, requiring the use of the hyperparameter tuner for optimization.

Accessibility is another key consideration. PtychoShelves is primarily implemented in MATLAB, which introduces licensing costs that can be a barrier to broader adoption, especially outside academia. PtyRAD and py4DSTEM are implemented in Python using only open-source dependencies, lowering the entry barrier for new users and fostering collaboration. While all three packages support GPU acceleration, PtyRAD uniquely supports multi-GPU acceleration (\cite{accelerate2022}), enabling reconstruction of large datasets that may not fit within a single GPU's memory and faster reconstruction times when using large batch sizes (SI Figure S2).

\begin{figure*}[ht]%
\centering
\includegraphics[width=0.9\linewidth]{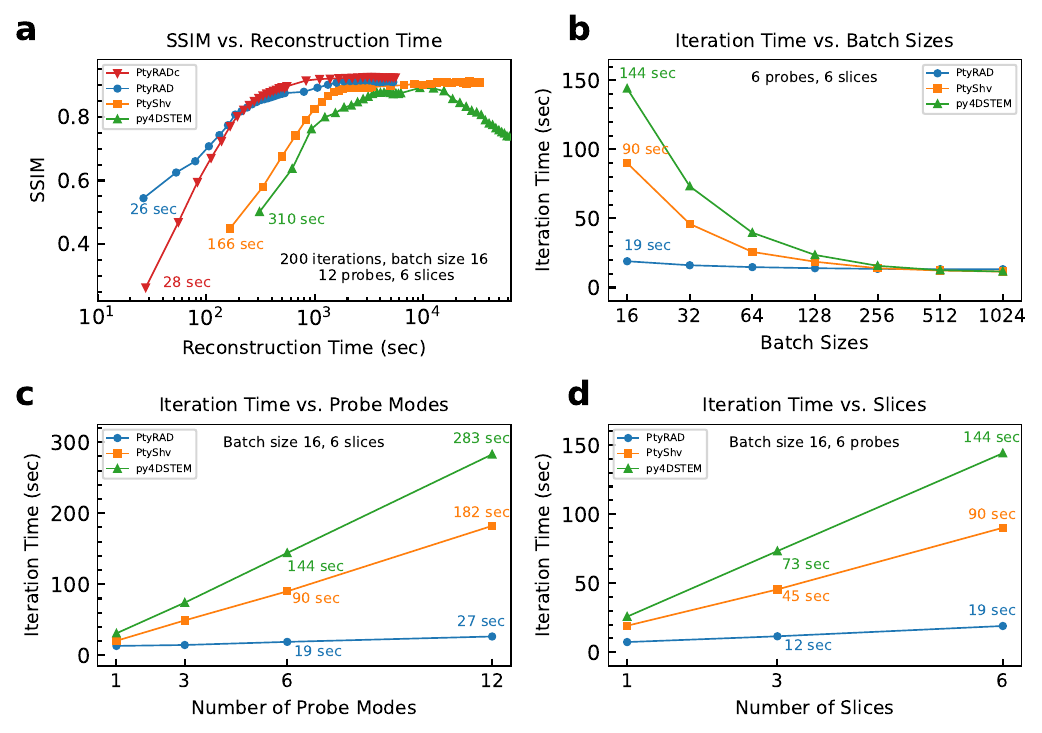}
\caption{Computation efficiency benchmark of PtyRAD against PtychoShelves (labeled as ``PtyShv'') and py4DSTEM using identical GPU-accelerated hardware and algorithm settings. (a) Reconstruction quality assessed by the structural similarity index measure (SSIM) comparing the reconstructed object phase with the ground truth using a simulated dataset. Higher SSIM corresponds to a better match with the ground truth. PtyRAD run with additional positivity constraint and sparsity is labeled as ``PtyRADc''. Scaling behavior of iteration times under different (b) batch sizes, (c) number of probe modes, and (d) number of object slices. The iteration time roughly scales inversely proportional with batch sizes for PtychoShelves and py4DSTEM, especially for small batch sizes, while PtyRAD only shows mild increase in the iteration time. Additionally, although the iteration time typically scales linearly with model complexities such as the number of probe modes and object slices, PtyRAD shows a much gentler slope due to PyTorch's optimized tensor operations and hardware acceleration, resulting the excellent computational efficiency of PtyRAD.}\label{fig_time}
\end{figure*}

\subsection{Computation speed and convergence}
To assess the computational performance of PtyRAD, we performed reconstructions with each package on identical datasets and using the same GPU hardware (a 20 GB MIG slice from an 80 GB NVIDIA A100). In this comparison, we choose Adam as the optimizer for PtyRAD, LSQML for PtychoShelves, and the ``gradient descent (GD)'' option for py4DSTEM. For each package, we recorded the iteration time excluding the package initialization and result-saving time to focus on the main computational work, and one iteration is defined as a full pass of all diffraction patterns in the dataset. We average all recorded iteration times for each package, ensuring that the warm-up effect from early iterations is negligible. Input files for each package to reproduce these reconstructions are provided in our Zenodo record, and we have made our best efforts to optimize the reconstruction settings for each package. See Supplementary Information for more benchmarking details.

To measure the convergence versus computation time of each package, we reconstruct a simulated dataset of twisted bilayer \ce{WSe2} and compare the structural similarity index measure (SSIM) (\cite{wang2004image}) of the reconstruction with the ground truth phase image at each iteration. This metric is preferable over the data error or loss function as it measures the perceptual similarity in terms of structure, brightness, and contrast of the reconstruction image. In addition, this is independent of the objective function used in the optimization and is not able to be arbitrarily lowered by overfitting.
The simulated dataset was generated using \texttt{abTEM} (\cite{madsen2021abtem}) with parameters comparable to the experimental tBL-\ce{WSe2} dataset, including phonon, partial coherence, and Poisson noise at a dose of $1 \times10^{6}$ e$^-$/\AA$^2$ (SI Table S4).
The ground truth 3D object phase is generated by first applying the strong phase approximation $t(\vec{x}) = \exp(i\sigma v_z(\vec{x})$ (\cite{Kirkland2010advanced}) to the \texttt{abTEM}-simulated 3D potential, and then taking the phase angle of the complex transmission function. The SSIM is calculated from the depth-summed 2D ground truth and reconstructed phase images. Note that we subtract the minimum value from each reconstructed phase image to remove any arbitrary global phase offset in the reconstruction, but phase differences are compared on an absolute scale, so failures to quantitatively reproduce phase contrast will be penalized.
The simulated dataset is available in our Zenodo record.

\begin{figure*}[ht]%
\centering
\includegraphics[width=0.75\linewidth]{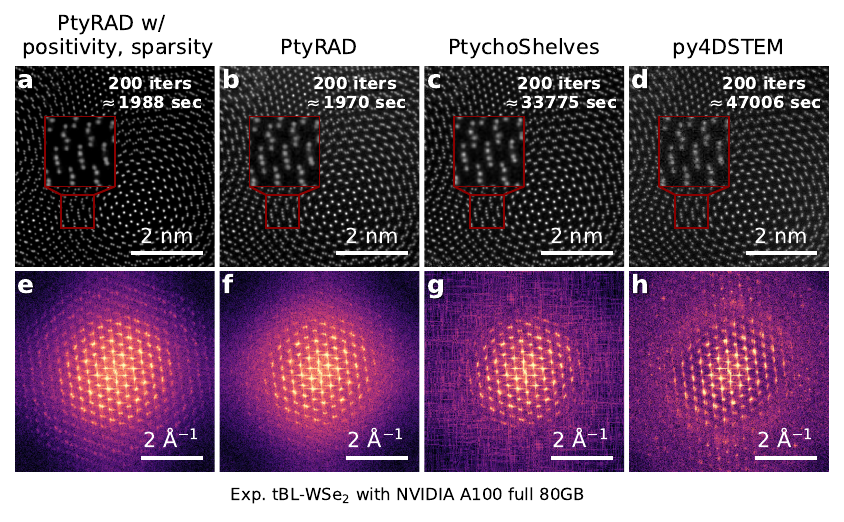}
\caption{Reconstructed phase images of the experimental twisted bilayer~\ce{WSe2} dataset after 200 iterations using (a) PtyRAD with positivity and sparsity regularization, (b) PtyRAD, (c) PtychoShelves, and (d) py4DSTEM and zoom-in insets. (e--h) Corresponding fast Fourier transform (FFT) power spectra of (a--d). Total reconstruction times taken for 12 probe modes, 6 object slices, batch size 16, and 200 iterations are labeled in the top right corners of each panel. PtyRAD achieves higher information transfer given the same number of iterations and a shorter reconstruction time.}\label{fig_convergence}
\end{figure*}

\begin{figure*}[ht]%
\centering
\includegraphics[width=0.9\linewidth]{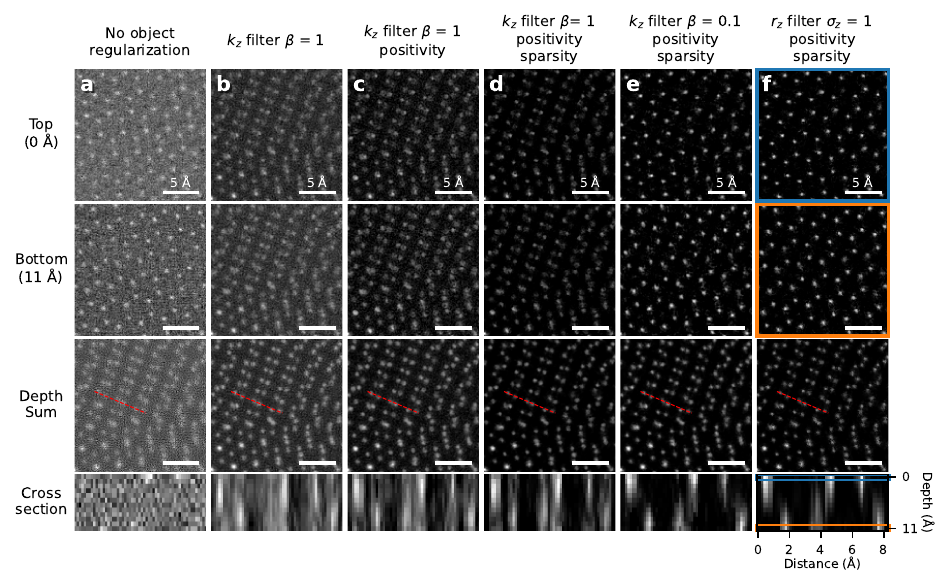}
\caption{Impact of object constraints and regularizations on multislice electron ptychography (MEP) tested with the experimental tBL-\ce{WSe2} dataset. The two layers are stacked in an anti-parallel configuration with an interlayer twist angle is 3 $^{\circ}$. We tested different constraint and regularization settings, including (a) no object regularization, (b) multislice $k_z$ filter $\beta = 1$, (c) $k_z$ filter $\beta = 1$ with object positivity constraint, (d) $k_z$ filter $\beta = 1$ with positivity and sparsity regularization, (e) $k_z$ filter $\beta = 0.1$ with positivity and sparsity, and (f) multislice $r_z$ filter $\sigma_z = 1$ with positivity and sparsity. The multislice $k_z$ and $r_z$ filters are necessary for stable MEP reconstructions, while the object positivity constraint and sparsity regularization improves the overall contrast and suppress noise. Each column displays the reconstructed top slice (0 \AA), bottom slice (11 \AA), the depth sum projection of all 12 slices, and a cross-sectional view along the red dashed lines labeled in the depth sum images. The blue and orange boxes highlight that the top and bottom slices reconstructed using the $r_z$ filter are well decoupled, showing reduced wrap-around artifact with a cleaner reconstruction quality compared to other regularization approaches. By properly combining the multislice $r_z$ filter, positivity, and sparsity regularizations, PtyRAD achieves much improved MEP reconstruction quality.}\label{fig_regularizations}
\end{figure*}

Figure~\ref{fig_time}a compares the progression of the SSIM with the accumulated reconstruction time over 200 iterations.
The corresponding final phase images and their FFTs are shown in SI Figure S3, while intermediate reconstructions and the ground truth potential are provided in SI Figure S4. 
Note that the accumulated reconstruction time is not the ``wall time''; it is defined as the previously mentioned average iteration time multiplied by the iteration number.
PtyRAD completes one iteration of reconstruction in 26~sec, a 6--12$\times$ speedup over the other packages (166 and 310~sec per iteration). 
This improvement stems primarily from the use of PyTorch--a highly efficient GPU computing framework--and a vectorized implementation of probe mode computations, which will be discussed later in more details.
We present two reconstructions using PtyRAD: the blue curve (``PtyRAD'') uses comparable settings to py4DSTEM and PtychoShelves, while the red curve (``PtyRADc'') includes our additional positivity and sparsity regularizations that will be further discussed in Figure~\ref{fig_convergence} and ~\ref{fig_regularizations}.
Notably, these constraints do not significantly affect iteration time, and both PtyRAD runs achieve comparable or higher SSIM than the other packages while requiring much less time.
It is important to note that due to the large space of possible algorithmic settings and the inherent variability between datasets, the convergence behavior of each package will vary in practice, and so these results may not apply universally across all applications.

Figures~\ref{fig_time}b--d illustrate how iteration time varies with key reconstruction parameters: the batch size (the number of diffraction patterns grouped together for processing per gradient update), the number of mixed-state probe modes, and the number of object slices, respectively\footnote{The timings here are measured using the experimental tBL-\ce{WSe2} dataset as in Figure~\ref{fig_exp_examples}c. This choice does not affect the comparison, as the timings are primarily dependent on the dataset size, rather than its content.}. 
The full table of iteration times is provided as SI Figure S5.

Figure~\ref{fig_time}b shows the iteration time as a function of the batch size, also known as grouping size or mini-batch size. 
The batch size does not change the total amount of computation required per iteration, but it sets an upper bound on the achievable level of parallelism, as patterns within a batch can be computed in parallel while individual batches must be processed sequentially.
Smaller batch sizes generally lead to longer iteration times due to lower GPU utilization, while increasing the batch size reduces the iteration time until the parallel computing pipeline of the GPU is fully utilized and no further improvement is observed. 
For very large batch sizes, with parallel computation of each pattern in the batch, the GPU memory will eventually become fully occupied---in this case, such large batch sizes cannot be processed by py4DSTEM or PtychoShelves.
PtyRAD supports processing larger batches by either spreading the workload across multiple GPUs (SI Figure S2), or by splitting the batch into sub-batches, which are computed serially and accumulate their gradients before applying the update step\footnote{A common technique in the machine learning community called ``gradient accumulation''.}. 
In practice, very large batch sizes are not commonly used with gradient descent ptychography as reconstruction quality tends to decrease significantly with increasing batch sizes (SI Figure S6).
Smaller batch sizes have been observed to yield better reconstruction quality, likely because they update the reconstruction much more frequently and introduce noisier gradients, which can help the optimizer to escape shallow local minima and improve convergence in non-convex problems. Conversely, reconstructions with larger batch sizes tend to stagnate or over-smooth fine features.
However, we anticipate that optimizations for large batches will be important as high-speed detectors with greater pixel count become available, as the memory required per batch increases sharply with the detector pixel count. 

Next, we investigate the performance impact of the model complexity, specifically the number of probe modes and object slices, for all three packages.
Figure~\ref{fig_time}c and d show that the iteration time scales linearly with increasing model complexity.
The linearity is expected because the iteration time is primarily dominated by the forward and backward passes, and the needed computation increases proportionally with the number of probe modes and object slices as outlined in \nameref{subsec_Physical model}.
Note that the curves in Figure~\ref{fig_time}c and d do not pass through the origin due to unavoidable GPU overheads\footnote{Common GPU overheads include data transfer, device synchronization, kernel launch, and other operations that incur time costs without performing actual computation.}.

It is also worth mentioning that both PtychoShelves and py4DSTEM process probe modes sequentially, whereas PtyRAD employs a fully vectorized approach that performs the forward and update steps for all probe modes in parallel. 
Since each probe mode contributes independently to the total diffraction intensity, PtyRAD vectorizes the computation by broadcasting the object tensor across probe modes as illustrated in  Figure~\ref{fig_overview}.
Parallelization of the probe mode calculations allows PtyRAD to more effectively fill the parallel computing pipeline of the GPU, leading to higher performance. 
In contrast, computation of each slice is necessarily serial and can not be vectorized for all three packages.

Additionally, the underlying GPU frameworks differ among these packages---PtyRAD leverages PyTorch, py4DSTEM utilizes CuPy, while PtychoShelves uses the MATLAB Parallel Computing toolbox.
These GPU frameworks play a critical role in performance as behind-the-scenes scheduling and dispatching of computational tasks to the GPU cores can significantly impact the GPU utilization.
Since each GPU framework employs different strategies for resource scheduling, kernel launching, and data caching, PtyRAD achieves the best performance at small batch sizes, while the advantage diminishes with increasing batch sizes as shown in Figure~\ref{fig_time}b and SI Figure S5.
We also report iteration times and GPU metrics for each package in SI Table S5, and observe that PtyRAD consistently achieves higher GPU and memory utilization rates, highlighting the impact of the underlying GPU framework on compute efficiency.

While PtyRAD shows faster iteration times across the range of parameters shown here, we note that the observed convergence speed in Figure~\ref{fig_time}a is determined by both the computation time and the improvement per iteration. 
Since both factors are highly sensitive to the chosen reconstruction parameters and properties of the dataset, the speed advantage of PtyRAD may not always translate into overall higher reconstruction quality when compared to other major packages. 

In our testing we have also found that the performance varies greatly with different GPU hardware, even when keeping the same reconstruction parameters. To understand which hardware specification has the most influence on reconstruction speed, we also benchmarked PtyRAD on different GPUs and find that the performance is roughly proportional to the GPU memory bandwidth under typical settings (SI Figure S7), and relatively less sensitive to the peak computing power measured in floating point operations per second (FLOPS).
This indicates that PtyRAD is primarily memory bandwidth-limited, meaning its throughput is ultimately constrained by how quickly data can be moved between GPU memory and compute cores. This limited bandwidth helps explain the linear scaling of PtyRAD observed in Figure~\ref{fig_time}c despite the probe modes are calculated in parallel. It also accounts for the saturation behavior seen in Figure~\ref{fig_time}b, where iteration times plateau at larger batch sizes across all three packages.

Figure~\ref{fig_convergence} compares the final reconstructed phase images, with their corresponding FFT power spectra, produced with each package from the experimental tBL-\ce{WSe2} dataset.
For all packages we used the same condition as in Figure~\ref{fig_time}a---12 incoherent probe modes, 6 object slices, a batch size of 16 patterns, and ran for 200 iterations. 
Note that a batch size of 16 yields the best reconstruction quality across packages under comparable conditions for this dataset (SI Figure S6).
The intermediate and final phase images are shown in SI Figure S8, while the reconstructed probe modes are presented in SI Figure S9.
The timings indicated on each panel for the total reconstruction time were measured using a full NVIDIA A100 80GB GPU.
Despite using the same number of iterations, PtyRAD achieves better contrast and higher information transfer while completing the reconstruction in 17--24$\times$ less time than the other packages given the test condition.



\subsection{Object constraints and regularizations}
Ptychography reconstructions are large-scale optimization problems involving millions of free parameters.
While ptychography experiments are designed to include multiple redundant measurements for each unknown parameter (i.e. to have a sufficient overlap ratio as discussed by \cite{bunk2008influence, edo2013sampling,da2015elementary}) and thus to overdetermine the solution (\cite{schloz2020overcoming, gilgenbachSamplingMetricsRobust2023}), the reconstruction is nevertheless ill-conditioned due to inherent ambiguities (\cite{rodenburg2019ptychography, li2016breaking}), imprecisely known experiment parameters, and noisy data.
Therefore, physical constraints and regularizations are indispensable for ptychographic reconstructions to ensure a physically meaningful solution.
Generally, regularization refers to penalty terms added to the loss function (e.g., sparsity regularization in Equation~\eqref{eqn_sparse}), whereas constraints correspond to operations that directly modify the solution (e.g., positivity constraint). 
While we follow this distinction, we use the term ``depth regularization'' to be consistent with existing literature, even though it acts more like a soft constraint.

Figure~\ref{fig_regularizations} demonstrates the impact of various constraint and regularization techniques on the reconstruction of the experimental tBL-\ce{WSe2} dataset. 
The reconstructions are all performed using 12 probe modes, 12 1-\AA\ object slices, and a batch size of 16, for 200 iterations.
The model contains $9,473,888$ optimizable parameters, including $9,047,904$ for the object, $393,216$ for the probe, and $32,768$ for the probe positions.
Meanwhile, the 4D-STEM dataset contains $128^4 = 268,435,456$ measured values, yielding an overdetermination ratio of $28.33$. 
Despite the seemingly sufficient overdetermination of the problem, with no regularization applied to the reconstruction, the phase image in Figure~\ref{fig_regularizations}a shows poor contrast and no clear separation between the two layers of the heterostructure along the depth direction. 

\subsubsection{Fourier-space depth regularization}
A fundamental challenge in MEP is the limited transfer of information along the beam propagation direction (\cite{terzoudis2023resolution}) combined with the need for thin object slices in order to correctly model multiple scattering of electrons. 
As a result, it is very difficult to reconstruct three-dimensional information without imposing regularizations on the object along the beam direction.
A common approach is the use of a low pass filter function in Fourier space, which downweights spatial frequencies along the $k_z$ direction for which there is minimal information transfer.
PtychoShelves, py4DSTEM, and PtyRAD all implement such $k_z$, or ``missing wedge'' filter (\cite{Wakonig2020PtychoShelves, chen2021electron}) with slight differences.
The missing wedge filter $W(\vec{k})$ is given by
\begin{equation}
W(\vec{k}) = 1 - \frac{2}{\pi}\tan^{-1} \left( \frac{\beta^2 |k_z|^2}{k_x^2 + k_y^2 + \epsilon} \right)
\end{equation}
where $\beta$ is the strength parameter, typically in the range 0.1--1, and $\epsilon = 10^{-3}$ is a small constant added for numerical stability. 
In PtychoShelves and PtyRAD, $W$ is further modified into $W_a$ by applying a lateral Gaussian blur with strength parameter $\alpha$, while py4DSTEM directly uses the original $W$ as the final filter function.
\begin{equation}
W_a(\vec{k}) = W \exp\left(-\alpha (k_x^2 + k_y^2)\right)
\end{equation}
The object function is then Fourier-filtered with the given filter function in reciprocal space, producing the modified object function $O'$.
\begin{equation}
O'(\vec{r}) = \mathcal{F}_{3D}^{-1}\left[W_a(\vec{k}) \cdot \mathcal{F}_{3D}\left[O(\vec{r})\right]\right]
\end{equation}
Figure~\ref{fig_regularizations}b shows that the contrast in the depth-summed phase image improves significantly by applying such $k_z$ filter with strength parameter $\beta=1$, but the separation between the top and bottom slices remains poor.

\subsubsection{Positivity and sparsity}
Arbitrary offsets of the object phase are also ambiguous, so we additionally apply an optional positivity constraint to ensure the object phase is strictly nonnegative in Figure~\ref{fig_regularizations}c.
By clipping negative phase values after each iteration, the reconstructed image  shows a higher contrast level and lower background intensity. 
We have also tested enforcing the positivity constraint by subtracting the minimum value at each iteration and found that clipping negative values provides much better reconstruction quality.

Since we expect phase images to contain bright, discrete atoms for atomic-resolution applications, for such applications we can also incorporate a sparsity-promoting regularization term (Equation~\eqref{eqn_sparse}) as shown in Figure~\ref{fig_regularizations}d. 
This additional sparsity term promotes ``atomicity'' (\cite{sayre1952squaring, van2012method, van2013general}) by adding a weighted L1 penalty term of the object phase into the loss function during reconstruction (Equation~\eqref{eqn_losses}).
The default weighting parameter of the sparsity term is set at $0.1$.
By simply suppressing small phase values, sparsity regularization improves visual quality with reduced background noises, enhanced contrast and layer separation, and extended information transfer (SI Figure S10). 

The sparsity promoting regularization appears to assist in deconvolution of the probe from the object by penalizing the ``ringing'' artifact around the atoms that occurs when the probe and object are not separated correctly. 
This also differs from simple intensity clipping, which can be used to artificially sharpen atomic resolution STEM images (\cite{yu2003artifacts}); in that case, intensity between atoms caused by the tails of the probe are artificially discarded.
In ptychography, the probe is deconvolved from the image, and the addition of this regularization term simply assists in correctly determining the probe function.
Therefore, positivity and sparsity regularization may enhance atom localization accuracy by suppressing background noise and reducing irregular phase variations around atomic columns, which often arise from probe--object intermixing.
However, care must be taken as the selective penalization of small values can potentially distort the relative phase contrast between different atomic species.
This is particularly concerning when imaging heterogeneous materials, as weaker atomic columns or isolated defects might be disproportionately affected or potentially eliminated by aggressive regularization.
Therefore, the regularization strength must be carefully selected and checked to prevent excessive suppression of weak signals and potential artifacts. 

\subsubsection{Regularization strength and artifacts}
The strength parameter $\beta$ of the $k_z$ filter presents a similar trade-off between the reconstruction stability and depth resolution.
This $k_z$ filter is designed to downweight noise in the ``missing wedge'' of k-space, which stabilizes the multislice reconstruction.
However, increasing $\beta$ also suppresses more $k_z$ information within the region, yielding a reconstruction with less details along the depth dimension.
As shown in Figure~\ref{fig_regularizations}e, reducing the $k_z$ filter strength from $\beta=1$ to 0.1 improves the separation of the top and bottom layers, as seen in both individual slices and cross-sectional views.
This improved layer separation primarily stems from preserving more $k_z$ information with a weaker filter, enabling recovery of finer structural features in depth.
At the same time, the weaker filtering also reduces the level of wrap-around artifacts, which can be particularly problematic in specimens lacking depth periodicity---such as twisted 2D materials or those with uncorrected crystal tilt, where the top and bottom slices do not perfectly align.
The wrap-around artifact ``transfers'' surfaces intensities to the other end of the volume via the periodic boundary condition inherent in Fourier space, effectively imprinting features from one surface onto the other. 
Although this effect is largely mitigated by reducing $\beta$, some residual wrap-around artifacts remain visible in the top and bottom slices of Figure~\ref{fig_regularizations}e.
In addition, we have observed that a $k_z$ filter with excessive regularization strength can introduce incorrect local intensity variations (SI Figure S11), highlighting the need for careful parameter selection and mindful interpretation.

Beyond reducing the strength parameter $\beta$, another common approach to mitigating the wrap-around artifact of the $k_z$ filter is to set the extent of the object array thicker than the actual specimen.
Conceptually, this introduces vacuum layers above and below the specimen.
However, this method presents several challenges in practice.
First, increasing the total number of slices adds computational overhead, further prolonging the already time-consuming MEP reconstructions.
Second, it introduces ambiguity in the specimen’s vertical position within the object array, as this position can be freely adjusted by altering the probe defocus during reconstruction.
Third, there is no guarantee that the reconstructed object will form well-defined vacuum layers.
In our observations, reconstructions with additional slices often produce specimens that either concentrate at one of the surfaces or become evenly "stretched" vertically to fill the entire object array, eliminating any vacuum regions.
While py4DSTEM offers an option to pad the object array before applying the $k_z$ filter and then crop it afterward, we find that this approach significantly reduces the reconstructed phase values at the surfaces for the tBL-\ce{WSe2} dataset.

\subsubsection{Real-space depth regularization}
In contrast to the Fourier-space $k_z$ filter, PtyRAD uniquely implements a real-space depth regularization, $r_z$ filter, by applying a 1D Gaussian blur along the object depth dimension. 
\begin{equation}
O'(\vec{r}) = O(\vec{r}) * G_{\sigma_z}(r_z)
\end{equation}

Here, $G_{\sigma_z}$ is the 1D Gaussian kernel with a standard deviation $\sigma_z$, and the modified object $O'$ is obtained by convolving the original object $O(\vec{r})$ with the Gaussian kernel $G_{\sigma_z}(r_z)$ in real space.

Unlike the $k_z$ filter (Figure~\ref{fig_regularizations}d--e), which operates in Fourier space and introduces the wrap-around artifact, the $r_z$ filter (Figure~\ref{fig_regularizations}f) stabilizes the multislice reconstruction without intermixing the top and bottom slices. 
Given that the interlayer distance of bulk \ce{WSe2} is approximately 6--7~\AA, our results (Figure~\ref{fig_regularizations}e--f) indicate a comparable depth discrimination as shown in their cross-sectional views.
To better estimate the depth resolution, we reconstructed with additional vacuum slices for a more accurate evaluation.
We measured the full width at half maximum (FWHM) of 34 single W atoms along the depth direction across different sample regions (SI Figure S12). 
This analysis yields a mean depth resolution of 7.5~\AA, with the best value reaching 6.6~\AA. 
These results highlight the importance of appropriate regularization strategies and careful parameter selection in achieving high-quality 3D reconstructions.

\begin{figure*}[ht]%
\centering
\includegraphics[width=0.9\linewidth]{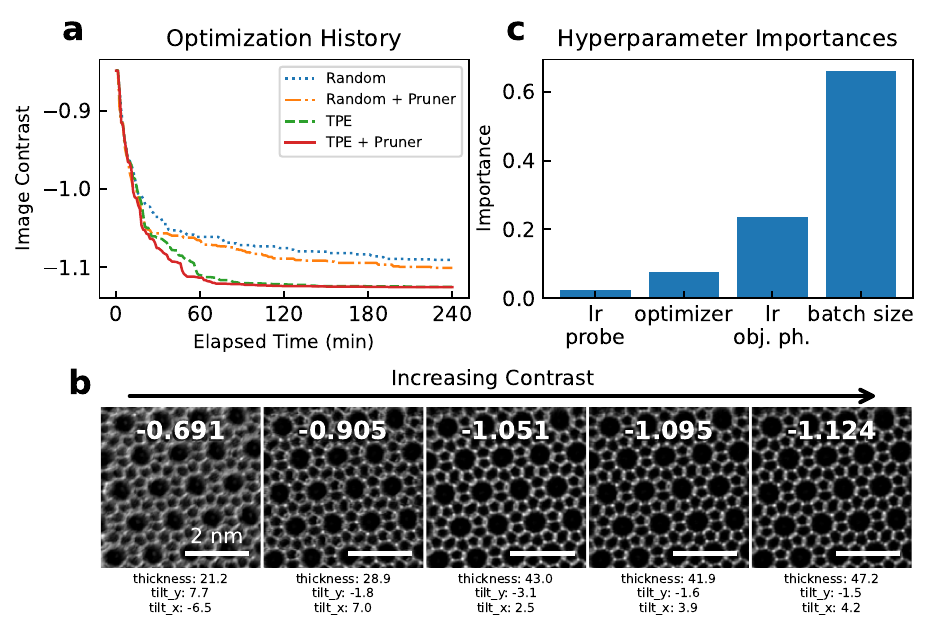}
\caption{Hyperparameter tuning using Bayesian optimization (BO) with pruning for ptychographic reconstruction on the experimental ZSM-5 dataset.
(a) Convergence of image contrast optimization using different search methods, including random sampling and BO with Tree-structured Parzen Estimator (TPE). Each method is evaluated with and without the Hyperband pruner. TPE combined with pruning demonstrates the fastest convergence and best performance. (b) Representative reconstructed phase images with image contrast values labeled on top, demonstrating the impact of hyperparameter tuning on image quality. The BO-suggested specimen thickness (\AA) and tilts (mrad) are labeled at the bottom. Note that a negative sign is added to the image contrast for the minimization direction. (c) Hyperparameter importances calculated using PED-ANOVA. Batch size is identified as the most influential hyperparameter, followed by the learning rate for the object phase (labeled as ``lr obj. ph.''), while the choice of optimizer and learning rate for the probe (denoted as ``lr probe'') show relatively low impact on the optimization objective.}\label{fig_BO}
\end{figure*}

\subsection{Hyperparameter tuning}
Beyond the millions of model parameters optimized through gradient descent, ptychographic reconstructions also depend on dozens of algorithmic and experimental parameters that critically influence reconstruction quality but are difficult to optimize directly via gradient-based methods.
These parameters are fixed during the reconstruction and are often referred to as hyperparameters (\cite{bergstra2011algorithms}).
To identify optimal choices for these hyperparameters, all three packages implement a Bayesian optimizer that runs separate trial reconstructions with different input values.
Both PtychoShelves (specifically the \texttt{fold\_slice} forks) and py4DSTEM utilize Bayesian optimization with Gaussian process (BO-GP) for their hyperparameter tuning.
PtyRAD uses Optuna (\cite{akiba2019optuna}), a widely-used open-source framework for distributed hyperparameter search, which supports a variety of probability models and optimization strategies. 
Note that the required accuracy of initial experimental parameters is strongly dependent on the dataset quality, especially the signal-to-noise ratio of the diffraction patterns.
For example, an initial guess within $\pm\ 5$ nm for the probe defocus is usually acceptable for dose-sufficient datasets like the tBL-\ce{WSe2} and PSO. 
In contrast, low dose datasets like MOSS-6 and ZSM-5 are much harder to recover the probe, so the initial defocus value becomes critical and requires hyperparameter tuning to optimize the probe defocus.

Figure~\ref{fig_BO} demonstrates the use of Bayesian optimization (BO) for tuning the algorithmic and experimental parameters for a reconstruction of the ZSM-5 dataset (\cite{zhangThreedimensionalInhomogeneityZeolite2023}) using a full NVIDIA A100 80GB GPU. 
Figure~\ref{fig_BO}a compares different hyperparameter search strategies based on their performance over time. 
The optimization targets two critical parameters affecting reconstruction quality: total sample thickness and crystal misorientation (tilts). 
While PtyRAD supports AD-based optimization of these parameters, they are typically optimized as hyperparameters in other packages, making them a natural benchmark for evaluating search strategies.
We compare random search with the Tree-structured Parzen Estimator (TPE) (\cite{bergstra2011algorithms}), a nonparametric BO algorithm that adaptively refines the sampling distribution by distinguishing between well-performing and poorly performing configurations.
Notably, TPE has been reported to outperform GP (\cite{bergstra2011algorithms}), and has been widely adopted in modern hyperparameter optimization frameworks developed by the machine learning community (\cite{watanabe2023tree}). 
TPE begins with random sampling for the first 10 trials to establish an initial estimate of the model response. 
Following these initial trials, the TPE method chooses superior hyperparameters which produce reconstructions with improved optimization objective. 

To improve the convergence speed of the BO optimization, Optuna supports pruning, which monitors the objective function improvement of each trial reconstruction during iteration and terminates underperforming trials early. 
Figure~\ref{fig_BO}a shows the convergence of both random and TPE search with and without use of the Hyperband (\cite{li2018hyperband}) pruner.
Each convergence curve is averaged over 20 independent runs, while each run is limited to a 4-hour time budget.
Each trial has a maximum of 20 iterations if not pruned.
To ensure fair comparisons, all strategies share the same 10 initial trials in each run, with random seeds assigned based on the run index. 
This guarantees that every strategy starts from an identical initialization before exploring further.
The TPE model with pruning (red curve) shows the fastest improvement of the image contrast and results in the most optimal solution in the allotted time.
On average, TPE with pruning explores 156 trials, of which 76 of them are pruned during the process.
Without pruning, only 125 trials were evaluated in the time limit, indicating the efficiency of pruning algorithm.
Although Figure~\ref{fig_BO}a suggests that on average, the image contrast converges within 120 minutes using the TPE algorithm with pruning, the actual time required naturally depends greatly on the initial accuracy of the experimental parameters and the signal-to-noise ratio of the data.

One critical challenge in hyperparameter tuning for ptychographic reconstruction is selecting an appropriate quality metric.
Conventional data error metrics computed from diffraction patterns do not always correlate with perceptual image quality (SI Figure S13) as overfitting can lead to solutions with lower error but nonphysical artifacts. 
SSIM provides an objective measure of reconstruction quality but requires a ground truth reference for comparison, making it irrelevant for experimental data. 
To address this, we use the image contrast, defined as the standard deviation ($\sigma$) divided by the mean intensity ($\mu$), as our objective metric for hyperparameter tuning.
This metric provides a simple, normalized measure of image variation that is linearly related to standard deviation, making it applicable for reconstructed phase images regardless of their absolute scales.
It has long been used for auto-focusing in electron microscopy (\cite{kirkland1990image, kirkland2018fine}) and is commonly referred to as the normalized variance (\cite{pattison2024beacon}), coefficient of variation (CV), normalized root-mean-square deviation (NRMSD), and relative standard deviation (RSD) (\cite{everitt2010cambridge}).

\begin{equation}
\text{Optimization Objective} = - \frac{\sigma(O_p)}{\mu(O_p)}
\end{equation}
Here, $O_p$ denotes the reconstructed object phase. For optimization purposes, we apply a negative sign to the contrast metric, ensuring a consistent minimization objective.
Figure~\ref{fig_BO}b shows the representative phase images reconstructed during the BO process, demonstrating a good correlation between image quality with the image contrast value. 

To better understand the influence of other general hyperparameter on the optimization objective, Figure~\ref{fig_BO}c calculates the importance of different hyperparameters, including batch size, learning rates, and choice of optimizers using PED-ANOVA (\cite{watanabe2023ped}), a statistical analysis that quantifies the relative importance of each hyperparameter by measuring its contribution to the total variance in optimization outcomes. 
The parameter search space for importance analysis is listed in SI Table S6.
The analysis reveals that batch size has the greatest impact, which is consistent with our previous observation (SI Figure S6). The learning rate for the object phase is also critical, as it sets the update step size during optimization and directly impacts the resulting image contrast.
In contrast, the choice of optimizer and learning rate for the probe have relatively low influence compared to batch size and object learning rate.
Although the importance values might vary with datasets, optimization setting, search spaces, and reconstruction configurations, this analysis provides valuable guideline for the overall workflow and is readily provided by Optuna and PtyRAD.

\section{Conclusion}\label{Conclusion}
In this paper, we presented PtyRAD, an open-source software framework for iterative ptychographic reconstructions.
By leveraging automatic differentiation and PyTorch’s optimized tensor operations, PtyRAD achieves up to a 24$\times$ speedup over existing packages. 
Our benchmarking results demonstrate that this performance improvement does not come at the expense of reconstruction quality, making PtyRAD a practical solution for high resolution and high throughput applications. 
Furthermore, we introduce a real-space depth regularization technique to mitigate wrap-around artifacts, which can be particularly useful for twisted 2D materials and vertical heterostructures.
By combining different regularization techniques, we demonstrate a mean depth resolution of 7.5\AA\ on an experimental tBL-\ce{WSe2} dataset with the best value approaching 6.6\AA.
In addition, PtyRAD’s integrated Bayesian optimization workflow streamlines hyperparameter selection, improving reconstruction robustness across diverse experimental conditions.

Looking forward, we aim to expand PtyRAD’s capabilities to support other 3D imaging modalities, including ptycho-tomography (\cite{ding2022three, romanov2024multi}) and tilt-coupled multislice electron ptychography (TCMEP) (\cite{dong2025sub}). 
Further computational performance can potentially be improved by alternative GPU frameworks with just-in-time (JIT) compilation capability, such as Jax (\cite{jax2018GitHub}). 
Most critically, we seek to establish more quantitative and reproducible ptychographic reconstruction strategies by exploring new metrics, optimization techniques, and regularization methods. 
By continuing to refine and extend PtyRAD, we hope to advance the field of ptychography and facilitate high-quality, interpretable reconstructions across a wide range of applications in electron ptychography and beyond.


\section{Data and code availability}
For review purposes, a minimal working example of the code and datasets are available at: \url{https://bit.ly/42HtQyD}

The PtyRAD package, along with the data, input parameter files, and code required to reproduce all figures, will be made publicly available at:
\begin{itemize}
    \item GitHub repository: \url{https://github.com/chiahao3/ptyrad/}
    \item Zenodo record: \url{https://doi.org/10.5281/zenodo.15273176}
\end{itemize}

\section{Supporting information}
To view supplementary material for this article, please visit [LINK].

\section{Competing interests}
No competing interest is declared.

\section{Acknowledgments}
The authors thank Dr. Yi Jiang, Dr. Xiangyu Yin, Dr. Amey Luktuke, and Dr. Ming Du for the support of X-ray compatibilities and valuable discussions. We also thank the members of the Muller group, particularly Dr. Guanxing Li, Harikrishnan KP, Lopa Bhatt, Noah Schnitzer, Clara Chung, Shake Karapetyan, Naomi Pieczulewski, Schuyler Zixiao Shi, and Zhaslan Baraissov for providing experimental datasets, sending feature requests, giving useful feedback during the development stage of PtyRAD. This project is supported by the Eric and Wendy Schmidt AI in Science Postdoctoral Fellowship, a program of Schmidt Sciences, LLC. The authors acknowledge the use of PARADIM computing resources under cooperative agreement number DMR-2039380.



\bibliographystyle{apalike}
\bibliography{00_reference}

\clearpage
\FloatBarrier  



\newpage
\includepdf[pages=-]{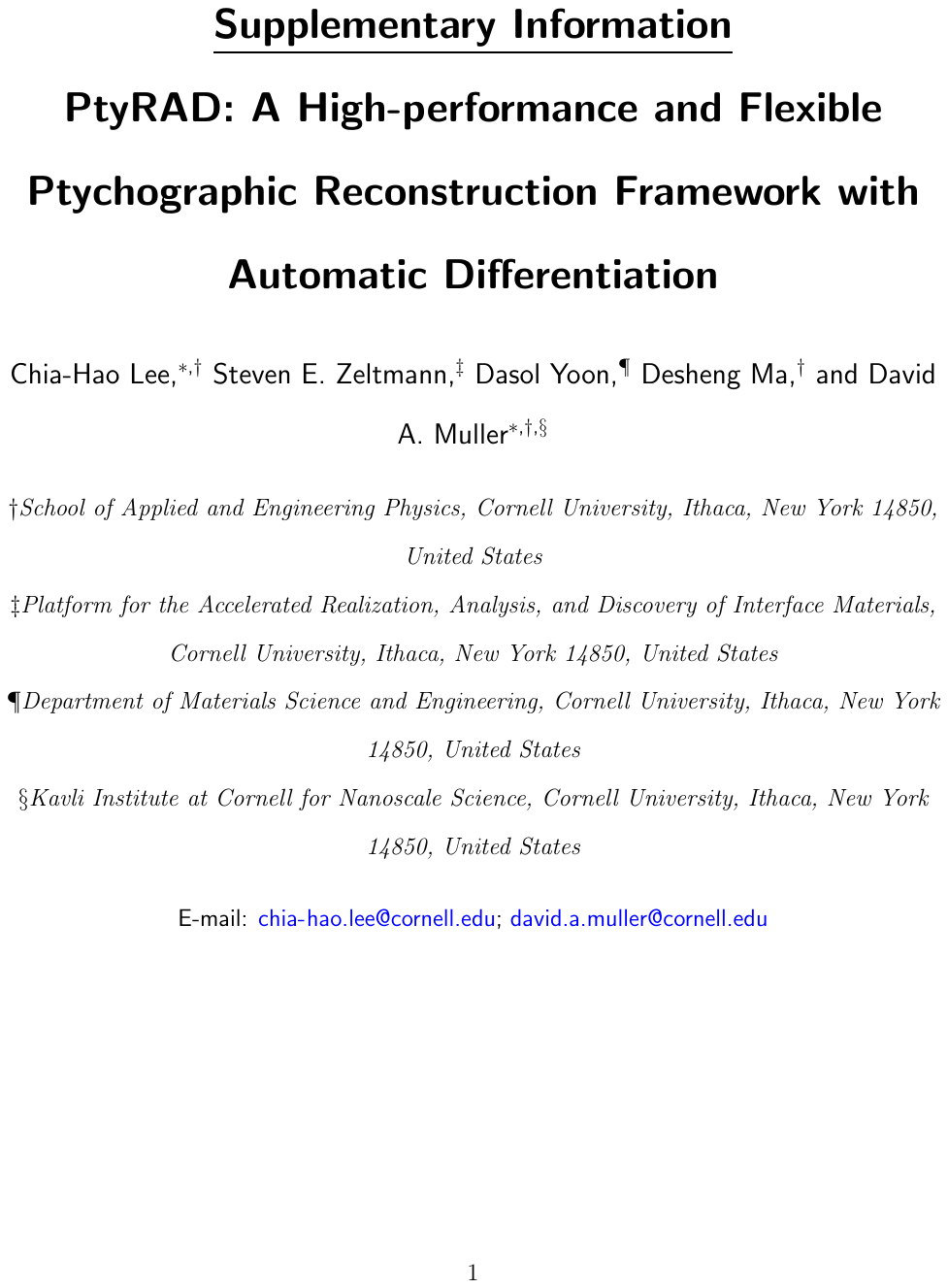}

\end{document}